# SND data acquisition system upgrade

**A.G.Bogdanchikov** [a,*], **V.P.Druzhinin** [a,b], **A.A.Korol** [a,b], **S.V.Koshuba** [a], **A.I.Tekutiev** [a], **Yu.V.Usov** [a]

[a] *Institute of Nuclear Physics,*
  *11, akademika Lavrentieva prospect, Novosibirsk 630090, Russia*

[b] *Novosibirsk State University,*
  *2 Pirogova Str., Novosibirsk 630090, Russia*
  *E-mail*: A.G.Bogdanchikov@inp.nsk.su

ABSTRACT: The data acquisition (DAQ) system of the SND detector successfully operated during four data-taking seasons (2010-2013) at the $e^+e^-$ collider VEPP-2000. Currently the collider is shut down for planned reconstruction, which is expected to increase the VEPP-2000 luminosity and data flow from the SND detector electronics by up to 10 times.

Since current DAQ system implementation (electronics and computer part) does not have enough reserve for selection of events in the new environment without compromising quality, there arose the need for the system upgrade.

Here we report on the major SND data acquisition system upgrade which includes developing new electronics for digitization and data transfer, complete redesign of the data network, increasing of the DAQ computer farm processing capacity and making the event building process concurrent. These measures will allow us to collect data flow from the most congested detector subsystems in parallel in contrast to the current situation. We would like to discuss also the possibility to implement full software trigger solution in the future.

---

[*] Corresponding author.

# Contents

## 1. SND data acquisition system

The data acquisition (DAQ) system of the SND detector (Fig. 1) successfully operated [1] [2] during four data-taking seasons (2010-2013) at the $e^+e^-$ collider VEPP-2000. Currently the collider is shut down for planned reconstruction, which is expected to increase the VEPP-2000 luminosity and data flow from the SND detector electronics by up to 10 times.

Since current DAQ system implementation (electronics and computer part) does not have enough reserve for selection of events in the new environment without compromising quality, here arose the need for the system upgrade.

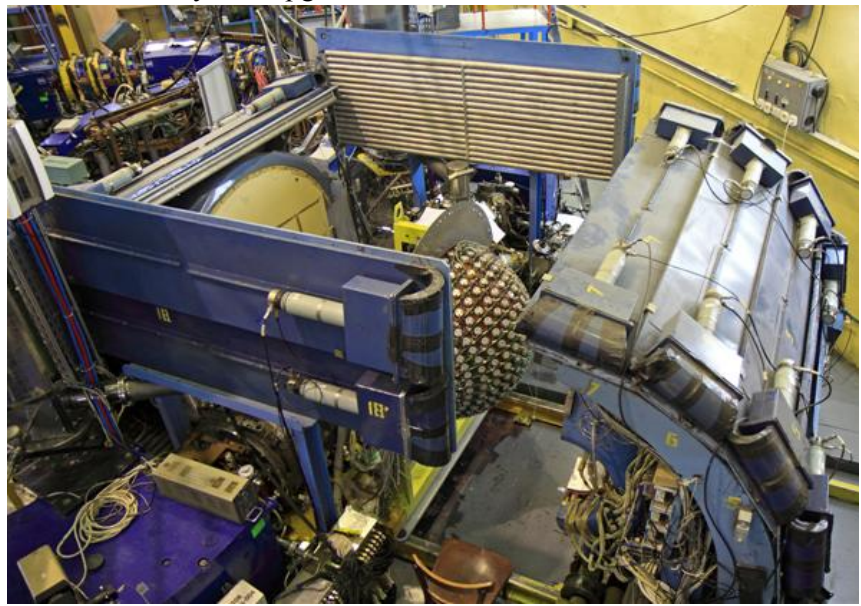

**Figure 1. SND detector**

## 1.1 Architecture

SND DAQ architecture is shown in Fig. 2. SND hardware trigger and channel statistics counting are implemented in the electronics in the CAMAC standard. After triggering digitized signals from the detector are read by the process READOUT from the electronics in the standard KLUKVA developed in BINP for the SND, KEDR, CMD-2 detectors. READOUT builds events from the collected data and stores them in the shared temporary buffer, where they become available for L3 processes. READOUT provides a small portion of received events for online visualization and operational control. L3 reconstructs and filters events for further offline processing.

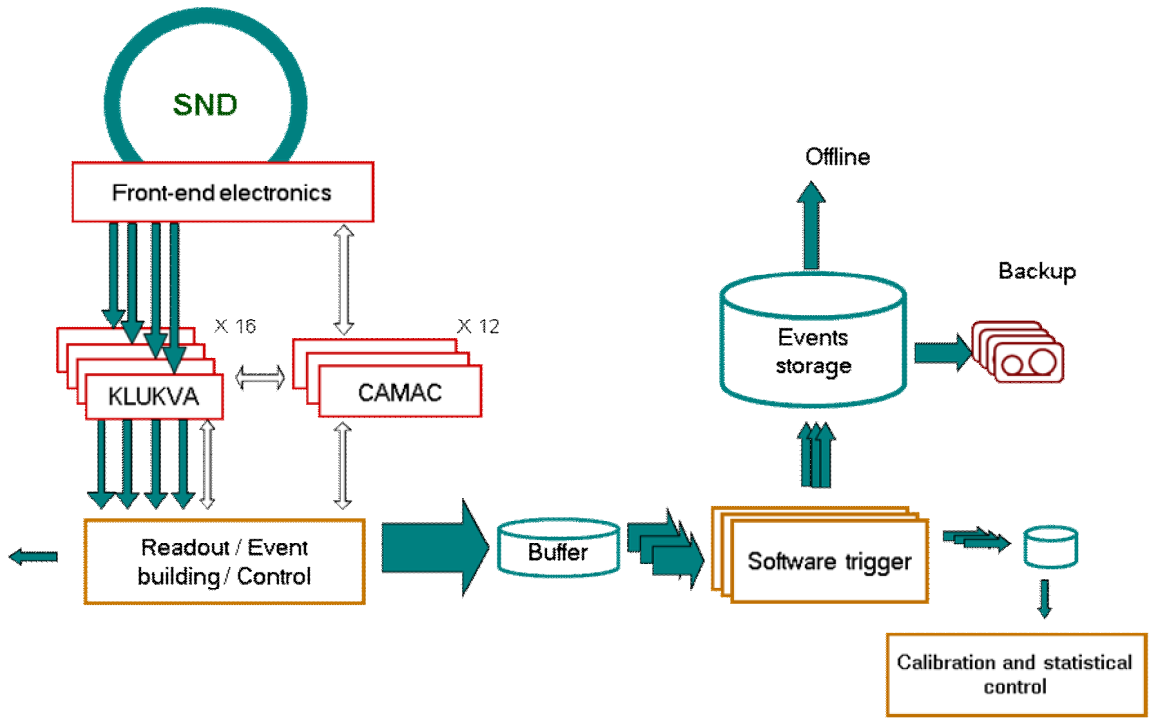

**Figure 2. Current SND DAQ architecture**

## 1.2 Dead time

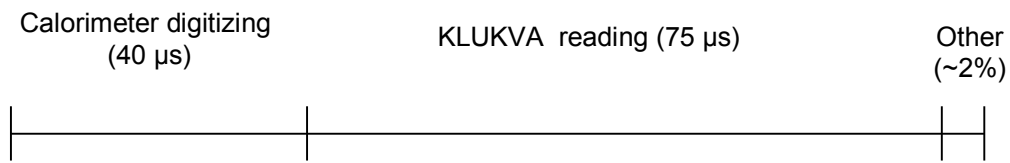

**Figure 3. SND DAQ dead time components**

SND DAQ dead time consists of the following parts (Fig. 3):

- Electronics digitizing dead time. Calorimeter signals digitizing takes up to 40 µs after triggering;
- KLUKVA reading dead time. Next event will not be triggered before all the data in all modules will be read through I/O processor in each KLUKVA crate. Crates communicate with computer in parallel, so the whole system dead time depends on the KLUKVA crate which generate the maximal amount of data. For SND experiment it is about 1.5 kilobyte, which corresponds to 75 µs/event of average dead time.
- Accelerator dead time. VEPP-2000 sends the block signal to SND DAQ system during the activities that may interfere with the normal operation of the detector.
- CAMAC reading dead time. Statistic counters and timer reading block periodically DAQ system for 1 ms.
- Buffer overflowing. The SND DAQ computer should have enough performance to handle the incoming data flows: network bandwidth, buffers capacity and computers power, otherwise the buffer overflows may increase the SND DAQ dead time.

The dependence on the lost event fraction from the event frequency (Fig. 4) is described quite accurately by the formula $f*\tau / (1+f*\tau)$, where f – spontaneous event frequency and τ – dead time per event. To preserve system performance at high event frequencies which are expected, the per-event dead time τ has to be reduced.

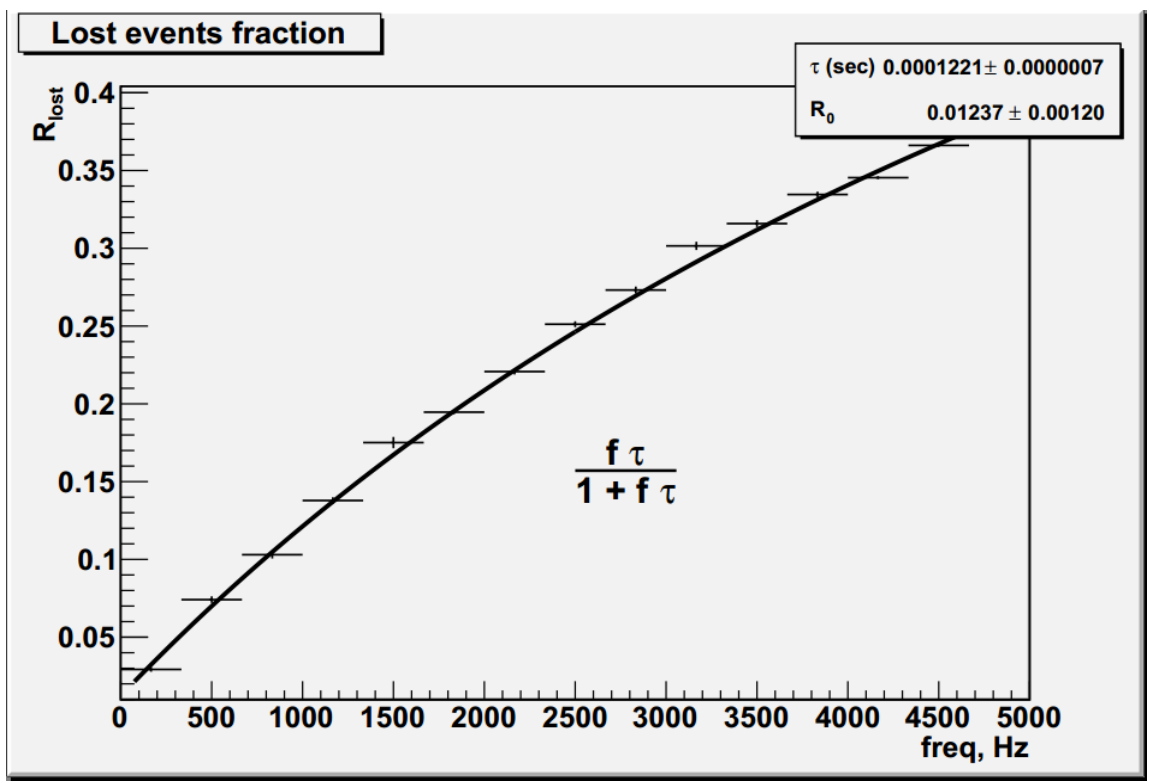

Figure 4. SND DAQ lost events fraction graph

## 2. SND DAQ upgrade

After the reconstruction of the VEPP-2000 accelerator complex, the load of the SND detector electronics is expected to increase by up to 10 times. For the treatment of the increased data flow the incremental improvements of the SND DAQ electronics are planned.

### 2.1 Digitising electronics

The new digitizing FPGA board A24F is developed. After receiving the trigger signal, the A24F board will digitize signals at 40 MHz frequency, implement zero suppression and send digitized data sets through the network directly to the event builder. As data are transmitted in parallel with the digitization, the board should not make no contribution to the SND DAQ dead time. Initially the board will be used for digitizing of the drift chamber cathode strips' signals. The removal of strip signal reading from the KLUKVA crates will reduce the maximum crate data size to 0.5 kilobytes, which corresponds to the 25 μs dead time. The electronics dead time will be reduced from 120 μs to 70 μs.

The next important step will be the replacement of the KLUKVA calorimeter digitizing boards to their network analogs. For this purpose the new calorimeter digitizing board is being developed:

- This board will perform not only signal digitizing, but also signal shaping, which will allow to exclude 140 shaping "F12" KLUKVA boards from SND DAQ.
- The data calorimeter will contain some additional temporal information important, in particular, for selection of e+ e- → n n̄ events.
- As the replacement of the calorimeter boards not only reduces the dead time to read, but also eliminates the dead time spent on digitizing, it will significantly improve the SND electronics loading capacity.

The digitizing boards from the other subsystems (drift chamber wires, Cherenkov counters, muon system) to be replaced by network analogs also.

## 2.2 Computer subsystem

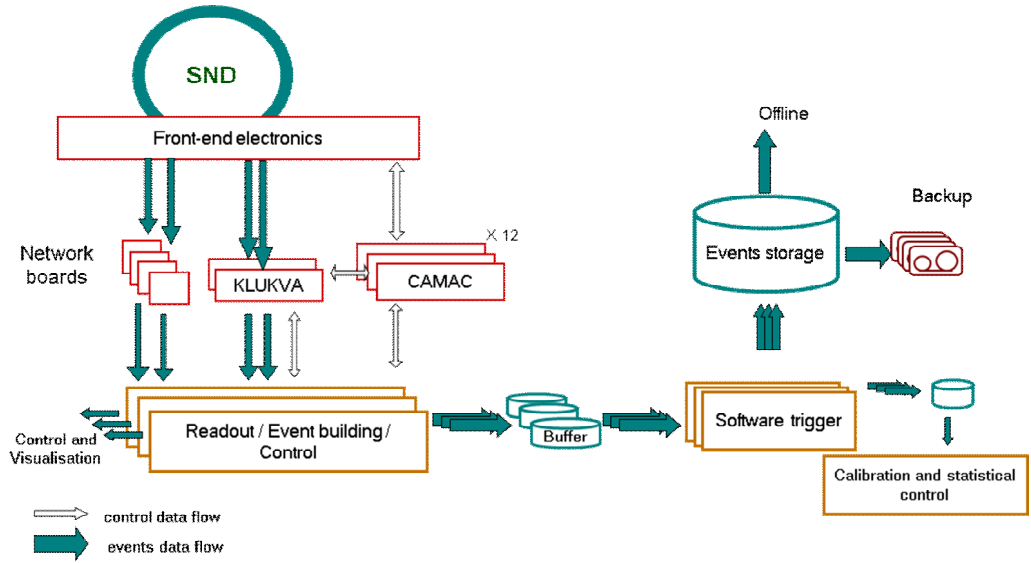

**Figure 5. Future SND DAQ architecture**

The SND DAQ computer subsystem will be upgraded (Fig. 5) in order process the data flows in the future experiments.

The Event Building (EB) process consumed about 60% of the CPU time (Intel Core i3 CPU 540 @ 3.07GHz) in the SND 2012-2013 experiment on beams. The increased event flow will require some additional power of EB, which is planned to achieve by the parallelization of event building procedure on the multiple CPU node. Each node will build sub-events from the specified digitization boards' subset. To make the data transfer from data builders more reliable, the network attached storage will be substituted with local buffers.

The online software trigger consumed about 30% of the online computer farm power in the SND 2012-2013 experiment. After upgrade the performance of the online farm will increase by about 10 times and should meet the DAQ performance requirements in the close future.

### 2.2.1 Fully software trigger

The most flexible and powerful event selection may be realized in the DAQ system with fully software trigger selection [3]. We consider to use this alternative in the future. The hardware trigger will no longer be used. After zeroes suppression, data will be passed and processed directly on the online computer farm.

The estimation of the load upper bound (12 Gb/s of sensible data plus 13 Gb/s of noise) allows to assert that fully software trigger DAQ will be the feasible task in the near future. In order to minimize the future expenses for development of new boards, all the boards that will be developed, should provide the ability to easily switch to fully software trigger mode.

## 2.3 Network

The events reading and building speed was up to 80 Mbps (7 KB at 1.4 kHz) at peak loads in the 2012-2013 SND experiment.

The network load will increase according to the next stages:

- Event flow will grow by a factor of 10, up to 800 Mbps (7 KB at 14 kHz).
- Network calorimeter electronics will sent the signal waveforms as digitizing arrays instead of measured amplitude values as it was before. The calorimeter arrays will increase the average event size to 9-10 KB, which is matching 1.2 Gbps event flow.
- Fully software trigger may produce up to 25 Gbps of events data at peak load. Additional methods for noise suppression and data compression may significantly reduce the rate.

As a result, the bandwidth of the SND network should be re-designed to handle 1.2 Gbps of event flow with hardware trigger and 25 Gbps if we consider the full software trigger.

## 3. Conclusions

Major SND data acquisition systems are under planned upgrade which includes developing new electronics for digitization and data transfer, redesign of the data network, increasing of the DAQ computer farm processing capacity and making the event building process concurrent. These measures will allow us to collect data flow from the most congested detector subsystems in parallel in contrast to the current situation. We also consider the possibility to implement full software trigger solution in the future.

## Acknowledgments

The work was supported by the Ministry of Education and Science of the Russian Federation, by the RF Presidential Grant for Scientific Schools NSh-2479.2014.2 and by the RFBR Grants No 13-02-00418-a, No 13-02-00375-a.